\documentclass[prl,twocolumn,superscriptaddress,floatfix,10pt]{revtex4-1}
\usepackage{graphicx,bm}
\usepackage{amsmath}
\usepackage{amsfonts}
\usepackage{amssymb}
\usepackage{braket}
\usepackage{dsfont}
\usepackage{bm}
\usepackage{eucal}

\usepackage{mathtools}

\usepackage{color}

\usepackage{xcolor}
\usepackage[colorlinks=true,urlcolor=blue,linkcolor=blue,citecolor=blue]{hyperref}
\usepackage{cleveref} 

\usepackage{comment}
\usepackage[normalem]{ulem}

\usepackage{tikz}
\usetikzlibrary{calc}
\usetikzlibrary{decorations.markings}
\usetikzlibrary{decorations.pathmorphing, arrows}
\usetikzlibrary{arrows.meta}

\usepackage[makeroom]{cancel}

\renewcommand{\vec}[1]{\boldsymbol{#1}}

\newcommand\eps{\epsilon}

\begin{document}
	\title{Divergent nonlinear response from quasiparticle interactions}
	\date{\today}
	\author{Michele Fava}
	\affiliation{Rudolf Peierls Centre for Theoretical Physics,  Clarendon Laboratory, Oxford OX1 3PU, UK}
    \affiliation{Philippe Meyer Institute, Physics Department, \'Ecole Normale Sup\'erieure (ENS), Universit\'e PSL, 24 rue Lhomond, F-75231 Paris, France}
	\author{Sarang Gopalakrishnan}
	\affiliation{Department of Physics, The Pennsylvania State University, University Park, PA 16802, USA}
	\affiliation{Department of Electrical and Computer Engineering, Princeton University, Princeton, New Jersey 08544, USA}
	\author{Romain Vasseur}
	\affiliation{Department of Physics, University of Massachusetts, Amherst, MA 01003, USA}	
	\author{Fabian Essler}
	\affiliation{Rudolf Peierls Centre for Theoretical Physics,  Clarendon Laboratory, Oxford OX1 3PU, UK}
	\author{S. A. Parameswaran}
	\affiliation{Rudolf Peierls Centre for Theoretical Physics,  Clarendon Laboratory, Oxford OX1 3PU, UK}

	\begin{abstract}
        We demonstrate that nonlinear response functions in many-body systems carry a sharp signature of interactions between gapped low-energy quasiparticles. Such interactions  are challenging to deduce from linear response measurements. The signature takes the form of a divergent-in-time contribution to the response -- linear in time in the case when quasiparticles propagate ballistically   -- that is absent for free bosonic excitations.  We give an intuitive semiclassical picture of  this singular behaviour, validated against exact results from a form-factor expansion of the Ising chain and tDMRG simulations in a non-integrable model --- the spin-1 AKLT chain. We comment on extensions of these results to more general settings, finite temperature, and higher dimensions.
    \end{abstract}
	
	\date{\today}
	
	\maketitle

    The response of a quantum many-body system  to external perturbations is 
	central to experimentally extracting information about its properties. In typical settings, such as transport and scattering measurements, the system is weakly perturbed out of its equilibrium state, and the leading  {linear} response~\cite{martin1968measurements, giuliani_vignale_2005} contribution can be related via the fluctuation-dissipation theorem to a two-point {equilibrium} correlation function. 	Consequently, linear response functions are often relatively straightforward to interpret.  For instance, at zero temperature ($T=0$), as long as the external perturbation can create single quasiparticle (QP) excitations on top of the ground state, their dispersion can be read off directly from momentum-resolved linear-response data. However, the simplicity that lends power to linear response often limits its utility in more complex situations.  For example, various distinct physical mechanisms can give rise to a broad frequency spectrum in linear response functions: e.g., selection rules requiring probe fields to excite multiple QPs, inhomogeneous broadening due to quenched disorder, and homogeneous broadening from QP decay. Differentiating among these using linear response data alone is a challenge.
	
    Often, nonlinear response functions give more direct insight into the nature of the low-energy excitation spectrum~\cite{mukamel1999principles}.  Intuitively, this is because they involve multi-time correlation functions that, suitably analysed, can disentangle different sources of broad spectra~\cite{PhysRevLett.122.257401}. Pump-probe experiments~\cite{Jepsen01a, PhysRevLett.87.237401} and two-dimensional coherent spectroscopy (2DCS)~\cite{lynch2010first, kuehn2011two, woerner2013ultrafast, lu2016nonlinear, mahmood2020observation, 2022arXiv220404203C} both extract nonlinear response functions. They have long been used in magnetic resonance and in optical experiments on chemical systems, usually in regimes where the  response reduces to that of individual atoms,  averaged over a suitable statistical ensemble. The extended many-body systems 
    usually encountered in solid-state materials or ultracold atomic gases  do not always admit a similarly simplified description. Developing techniques to compute nonlinear response functions in such systems is thus an important goal~\cite{2020arXiv200107839L, PhysRevLett.124.117205, 2021arXiv210106081K, 2021arXiv211010496S, PhysRevB.103.195133, PhysRevX.11.041006, PhysRevX.11.041007, PhysRevB.102.165137, Watanabe2020, 2021arXiv210511378T, 2022arXiv220408365M, PhysRevB.94.245121, PhysRevX.11.031035, arxiv.2209.00720, https://doi.org/10.48550/arxiv.2208.12817, https://doi.org/10.48550/arxiv.2209.14070, PhysRevLett.128.076801}, made more pressing as experiments begin to probe such regimes. Apart from situations that reduce to an ensemble of few-body systems~\cite{PhysRevLett.122.257401, PhysRevLett.125.237601, PhysRevResearch.3.013254}, most controlled results have been obtained for 
    free theories~\cite{PhysRevLett.115.216806, PhysRevLett.122.257401, PhysRevB.99.045121, Jo_o_2019, 2021arXiv210104136P}, or exactly-solvable models~\cite{10.21468/SciPostPhys.5.5.054, Doyon2019, 10.21468/SciPostPhys.8.1.007, PhysRevB.103.L201120, PhysRevB.104.205116,doi:10.1073/pnas.2106945118}. There is thus a need for qualitative insights into universal aspects of nonlinear response outside these settings.

    Here, we offer such a qualitative perspective:  a semiclassical theory of nonlinear response. We  focus on the simplest  non-trivial  systems,  whose low-energy spectrum consists of gapped  QPs and where the perturbing field can excite single QPs. For ballistic QPs in one dimension ($d=1$) and at $T=0$ we find that the $q=0$ third-order response functions diverge linearly in the time interval between distinct applications of the external field,  with a strength  set by the inter-QP scattering phase shifts, and  a  scaling function  specific to the nonlinear  protocol. This richness is to be contrasted with $q=0$ {linear response} for such systems: a  delta-function peak at the gap frequency, 
    related to the stability of QPs, and is nonzero even for free bosonic QPs with no scattering.
    
    Apart from enjoying the simplifying features of  ballistic $d=1$ QPs, the primary example we consider below -- the transverse-field Ising chain -- is also integrable. As is well known, it maps to a theory of free fermionic QPs, whose  statistics enforce a scattering phase shift of $-1$ leading to singular  nonlinear response.
    We benchmark semiclassical analysis for the Ising model against exact results using form-factor expansions, detailed in upcoming work~\cite{the_other}.
    While the form-factor approach is applicable to a subset of integrable models, the semiclassical approach can be generalized more broadly. 
    Within the same framework we can also treat non-integrable systems, as long as they feature  stable, gapped QPs in some range of momentum. To confirm the validity of the semiclassical approach in this context we perform tDMRG simulations in the (non-integrable) AKLT spin-$1$ chain~\cite{PhysRevLett.59.799, Affleck1988}.
    Furthermore, with only slightly more work, the semiclassical approach can be generalized to treat  [low] finite $T$. We also conjecture  that many features persist in $d>1$. Relaxing the restriction to $q=0$ to allow momentum resolution permits direct extraction of the QP scattering matrix by combining linear and nonlinear response data. Our work thus promises an intuitive way to compute and understand nonlinear responses in a variety of quantum many-body systems.

\textit{Setup.---}%
    As noted above, we  initially focus on the transverse field Ising chain,  with Hamiltonian 
    $H = -J \sum_{j=0}^{L-1} \left(\sigma^z_j \sigma^z_{j+1} + g \sigma^x_j \right)$
    In particular, we work in the paramagnetic phase ($g>1$) and consider $q=0$  perturbations coupling to the order parameter $M=\sum_j \sigma_j^z$ (recall that only such  `Ising-odd' operators can excite single QPs above the ground state). Extensions to $q\neq 0$ are straightforward and will be reported in detail in \cite{the_other}.

    The model is exactly solvable by means of a Jordan-Wigner transformation~\cite{*[{See, e.g. Appendix A of }] [{.}] Calabrese_2012}
    that maps $H$ 
    to a quadratic fermion problem. This yields a QP dispersion relation $\eps(k) = 2J \sqrt{1+g^2-2 g \cos(k)}$, with a gap $\Delta=\eps(0)$. However, since $\sigma^z$ maps  to a non-local (string-like) operator, exact response functions involving $M$ cannot be easily computed  using Wick's theorem, and instead require a form-factor expansion using techniques developed in Refs~\cite{PhysRevB.78.100403,EK09,pozsgay2008form,pozsgay2010form,PhysRevLett.106.227203,Calabrese_2012,Schuricht_2012,PhysRevLett.109.247206,10.21468/SciPostPhys.9.3.033}. Similarly, any local spin operator that can create single fermionic QPs must be nonlocal in the fermion basis. Consequently, their nonlinear response is sharply distinct from that of fermion-local spin observables~\cite{PhysRevLett.122.257401}, that only excite even numbers of QPs.
    
    We first study the pump-probe signal,
    \begin{align}
    \label{eq:non-pert-PP}
        \Xi_{\text{PP}} &= -\frac{i}{L} \braket{0|e^{i\mu M(0)} [M( t_1+ t_2), M( t_1)] e^{-i\mu M(0)}|0} \nonumber \\
        &\qquad +\frac{i}{L} \braket{0|[M( t_2), M(0)]|0},
    \end{align}
    which can be viewed as the difference in the {linear} response (measured between times $ t_1,  t_1+ t_2$) computed in two states: the original QP vacuum $\ket{0}$, and a `pumped' state $e^{-i\mu M(0)}\ket{0}$ obtained by perturbing the QP vacuum  at $ t=0$ with a `kick' of strength $\mu$ coupling to $M$~\cite{suppmat}.

    In the response regime where the pump only weakly perturbs the system away from equilibrium, we can expand $\Xi_{\text{PP}}$  in $\mu$ and evaluate the resulting correlators in equilibrium: odd powers vanish by  symmetry, so $\Xi_{\text{PP}}= \mu^2\chi^{(3)}_{\text{PP}}+O(\mu^4)$.
    The superscript denotes a third-order response, which we split into pieces divergent  and convergent in time, $\chi^{(3)}_{\text{PP}}= \chi^{(3)}_{\text{PP};d} +\chi^{(3)}_{\text{PP};c}$, where the former
    \begin{align}\label{eq:chiPP3d}
        \chi_{\text{PP};d}^{(3)} 
        &= \frac{2}{L}\Im\underbracket{\bra{0}M e^{iH( t_1+ t_2)}}_{\text{bra}}
        {\underbracket{M e^{-i H  t_2} M e^{-i H  t_1} M\ket{0}_C}_{\text{ket}}},
    \end{align}
    is our focus.    Only the connected correlator (denoted by the subscript $C$)  contributes to the response, as required by causality.
    Both from the Heisenberg picture in \eqref{eq:non-pert-PP} and the Schr\"odinger one \eqref{eq:chiPP3d}, it is evident the correlators are not time-ordered.
    It is convenient to also distinguish the `ket' and `bra' sides of \eqref{eq:chiPP3d}. 
    Formally these correspond to forward and backward branches of the Keldysh time contour, which runs from $ t=0$ to $ t= t_1+ t_2$ and back. At $ t=0$, the operator $M$ acts on bra and ket sides, whereas at time $ t= t_1$ it acts solely on the ket side. Both sides are then evolved up to time $ t= t_1+ t_2$, whereupon $M$ is measured. This sequence {\it is} path-ordered on the Keldysh contour.
    We adopt the standard nomenclature where  $n$\textsuperscript{th} order  response functions involve $n$ external perturbations and $n+1$ operators; the extra operator corresponds to the measured observable.
    [Note that $\chi^{(3)}_{\text{PP}}\varpropto \mu^2$, but appears at third order when expanding in terms of {\it all}  external fields, as an extra perturbation is required to extract the linear response function in Eq.~\eqref{eq:non-pert-PP}, cf.~\cite{suppmat}.]

\begin{figure}[ht]
	\centering
        \includegraphics[width=\linewidth]{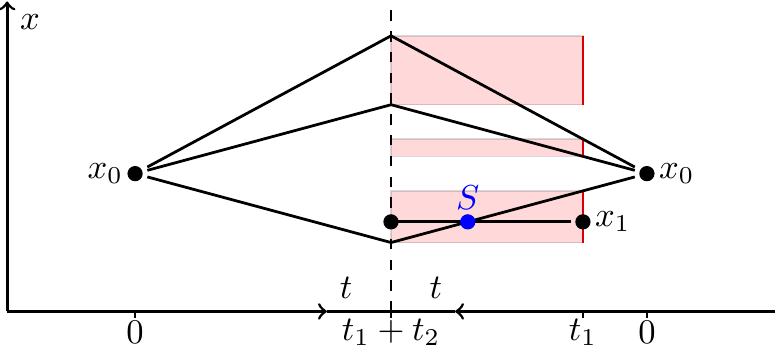}
	    \caption{
		    Cartoon of processes contributing to the late-time divergence of the third-order response $\chi^{(3)}_{PP;d}\sim\langle M(0)  M( t_1+ t_2) M( t_1) M(0)\rangle$. A dashed line, corresponding to time  $ t= t_1+ t_2$ separates the bra (left) and ket (right) sides of the time evolution; time increases towards this line.
            Solid lines denote QP worldlines, and circles denote the action of the operator $M$. When two lines cross, the amplitude for the diagram is multiplied by $S=-1$. For a fixed $x_0$, the red segments indicate the set of $x_1$ giving rise to a scattering-connected contribution. The length of the red set is proportional to the overall timescale, leading to the linear divergence in \eqref{eq:pump-probe-scaling}.
    	}
	    \label{fig:fig1}
    \end{figure}

\begin{figure}
        \centering
        \includegraphics[width=\linewidth]{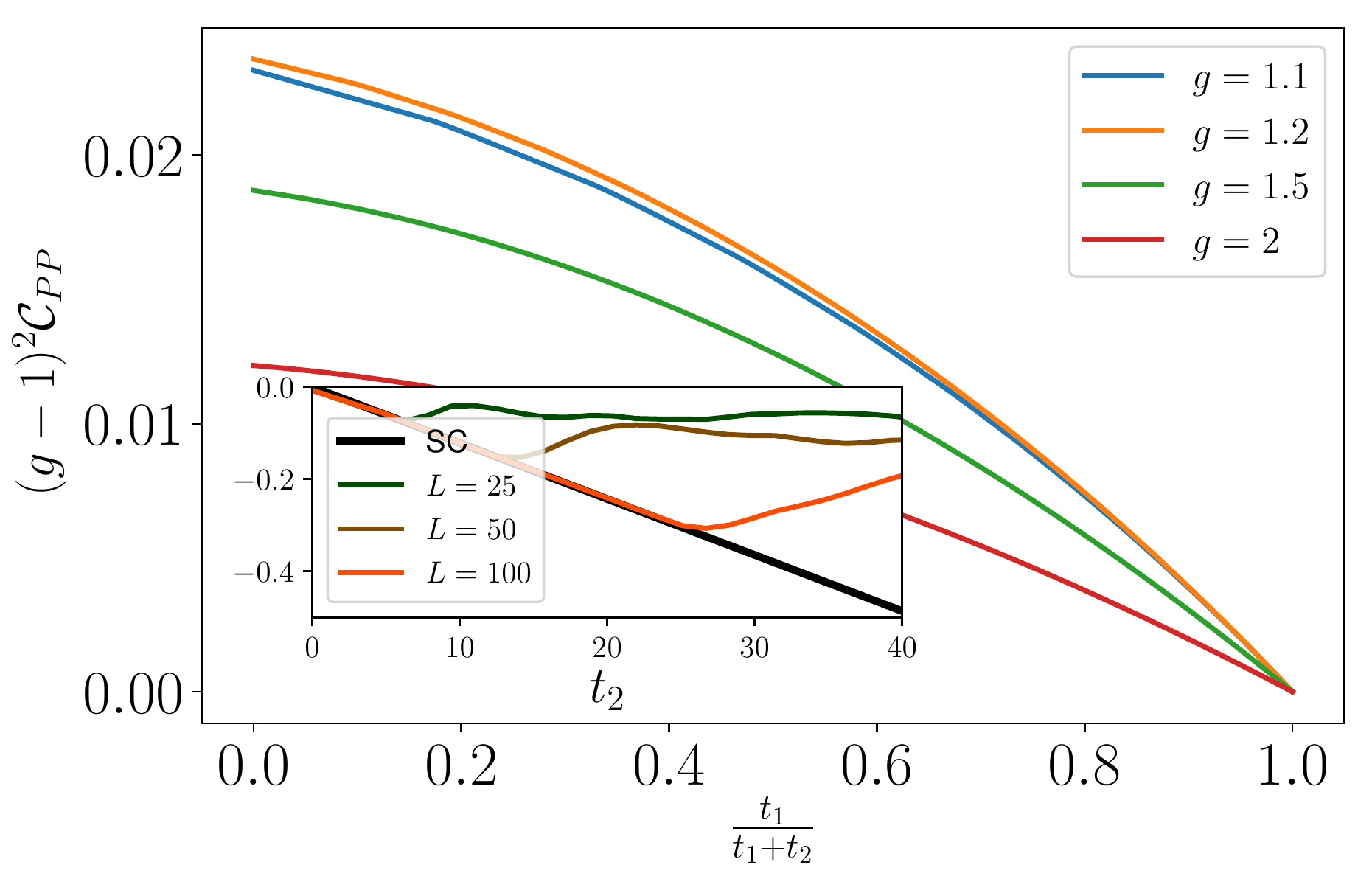}
        \caption{Scaling function $\mathcal{C}_{\text{PP}}(\frac{ t_1}{ t_1+ t_2})$ as defined in \eqref{eq:pump-probe-scaling} for various values of the transverse field $g$ and $J=1$. For graphical convenience, $\mathcal{C}_{\text{PP}}$ has been rescaled by $(g-1)^2$. Inset: $\chi^{(3),[3,4,3]}_{PP} / (2\sin(\Delta  t_2))$ for $ t_1=0$, $g=2$, and $J=1$. Various coloured curves denote the exact results obtained by numerically summing the form factors in an $L$-sites chain. The black line gives the linear scaling due to $\mathcal{C}_{PP,3}$ computed semiclassically, cf. \eqref{eq:mathcal-C-PP-n}.}
        \label{fig:fig2}
    \end{figure}

\textit{Long-time divergences and non-perturbative effects.---}
Our main result is that $\chi_{\text{PP};d}^{(3)}$ diverges at late times,
    \begin{equation}
    \label{eq:pump-probe-scaling}
        \chi_{\text{PP};d}^{(3)} \simeq 2 \sin(\Delta t_2) ( t_1+ t_2) \mathcal{C}_{\text{PP}}\left( \frac{ t_2}{ t_1+ t_2} \right) 
    \end{equation}
    with a scaling function $\mathcal{C}_{\text{PP}}$ whose numerical behaviour is shown in Fig.~\ref{fig:fig2}.
    This divergence can be given a simple semiclassical interpretation, involving ballistic propagation of quasiparticles and their scattering (Fig.~\ref{fig:fig1}).
    
    Before detailing the semiclassical analysis, we argue that we expect $\chi_{\text{PP}}^{(3)}$  to diverge on general grounds. Even for arbitrarily small  perturbation strength $\mu$, $\Xi_{\text{PP}}$ will saturate to an $O(1)$ value independent of $\mu$ at late times. This is most easily seen for sufficiently large $ t_1$, such that the system effectively thermalizes after the initial kick.  The perturbed state $e^{-i\mu M(0)}\ket{0}$ is then effectively at a finite (but small) $T$. [In the integrable case, it can  be approximated by a generalized Gibbs ensemble (GGE), but this is not essential to the analysis.]
    The first line of \eqref{eq:non-pert-PP} is then approximately a linear-response function in a finite-$T$ state, which on general grounds is expected to decay exponentially with dephasing rate $\gamma_\mu$,
    \begin{equation}\label{eq:PPcorr}
        \langle e^{i\mu M(0)} [M( t_1+ t_2), M( t_1)] e^{-i\mu M(0)}\rangle\sim e^{-\gamma_\mu  t_2},
    \end{equation}
    intuitively due to stochastic scattering events with the QPs created by $e^{-i\mu M(0)}$~\cite{sachdev1997low}. 
    Therefore, at long times, the effect of the perturbation become $O(\mu^0)$. A natural possibility is that $\gamma_\mu\varpropto \mu^2$, suggesting that $\chi^{(3)}_\text{PP}$ diverges linearly in $t_2$ whenever we have a stable QP excitation at zero momentum.
    The full nonlinear response thus initially shows a linear divergence, probed in the response limit, which eventually saturates at late times $ t\gg1/\mu^2$.
        
	\textit{Semiclassical picture.---}%
	The scaling form~\eqref{eq:pump-probe-scaling} and the scaling function $\mathcal{C}_{\text{PP}}$ can be quantitatively understood from a simple semiclassical picture inspired by the seminal approach of Refs.~\cite{sachdev1997low, SACHDEV1996576} (see also Refs.~\cite{PhysRevB.84.165117, Blass_2012, Evangelisti_2013, PhysRevE.93.062101}).  Our basic objects are wave-packet (WP) states $|r,k\rangle$ of QPs, which we will think of as having approximately well-defined positions $r$ and momenta $k$ in the sense that the effects of the dispersion of the wave packets will be sub-leading. Multi-WP states $\ket{\vec{r}; \vec{k}}$ are obtained as tensor products of single-WP states and by locality of the Hamiltonian (approximately) have simple dynamics as long as the WPs are spatially well separated. By construction $n$-WP states are in one-to-one correspondence with scattering states of $n$ QPs. On the ket side, the action of the operator $M(0)$ will after a sufficiently long time result in ``intermediate'' $n$-WP states (with $n$ odd), where the individual WPs approximately originate from the same point $x_0$ (which is integrated over) and whose momenta  sum to $0$. Each WP (approximately) propagates ballistically with its group velocity $v(k)=\eps'(k)$, i.e. $e^{-i H t} M(0)\ket{0}$  is approximately a superposition of states of the form $\int dx_0\, \ket{\vec{r}; \vec{k}}$
    with $\vec{r}$ specified by $r_j=x_0 + v(k_j) t$ and $\vec{k}$ such that $\sum_j k_j\approx0$, but otherwise arbitrary.
    We start by considering processes where all WPs proceed undisturbed to time $ t_1+ t_2$, whereupon they annihilate with the $n$ WPs produced by $M(0)$ on the bra side.  Note that, in order to have a non-negligible overlap between the bra and the ket, the momenta $\vec{k}$ on the two sides must approximately coincide, as well as the position $x_0$ at which the WP shower is created (see Fig.~\ref{fig:fig1}). In the Ising model the amplitude associated with creating and annihilating a shower of $n$ WPs with a given set of momenta $\vec{k}$ is $\left|F_n(\vec{k})\right|^2\frac{d^n \vec{k}}{(2\pi)^{n-1}} \delta\left(\sum_{j=1}^n k_j\right)$ for every initial position $x_0$. Here $F_n(\vec{k})=\braket{\vec{k}|\sigma^z_0|0}$ is the so-called $n$-QP form factor on top of the ground state, whose  precise value is unnecessary to proceed with the semiclassical calculation.
    [Note that in principle $F_n(\vec{k})$ and $v(k)$ can be numerically computed for non-integrable systems using MPS~\cite{10.21468/SciPostPhysLectNotes.7}.]

    In the processes of interest, $M( t_1)$ creates a single $q=0$ WP at  $x_1$, which is spatially well separated from the position at time $t_1$ of the $n$ WPs produced by the action of $M(0)$ on $|0\rangle$. The WP created by $M( t_1)$ is then annihilated by $M( t_1+ t_2)$, giving rise to an amplitude $|{F}_1(0)|^2 e^{-i\Delta t_2}$.	
    Naively, integrating over the arbitrary positions $x_0$ and $x_1$ generates a contribution to $\chi^{(3)}_{\text{PP}}$ that diverges proportionally to $L$ in the thermodynamic limit. However, the contribution of processes in which the WP trajectories do not cross is simply equal to the product of two-point functions $L^{-1}\langle M(0) M(0) \rangle \langle M( t_1+ t_2) M( t_1) \rangle$, and therefore the extensive spatial divergence cancels when we subtract the disconnected contributions. 
	
    In contrast, each time a pair of WP trajectories cross, the amplitude for the process picks up a factor of the scattering matrix $S$. In the Ising model, this simply encodes the fermionic statistics of QPs, i.e. $S=-1$. Therefore, a process involving an odd number of scattering events  (cf. Fig.\ref{fig:fig1}) does not cancel  against disconnected contributions; we refer to 
    processes which are connected only because of such events as \textit{scattering-connected}.
    After subtracting  the disconnected component we obtain a factor $S-1$, evaluating to $-2$ in the Ising case. Finally, we must integrate over all possible $x_1$ that produce  a scattering-connected process. [The integration over $x_0$ will produce a factor $L$, cancelling the $L^{-1}$  in the definition of $\Xi_{\text{PP}}^{(3)}$.] One can verify that, for a given set of momenta $\vec{k}$ of the WPs produced by the pump, the range of $x_1$ leading to scattering-connected processes has length $v_{\text{PP}}( t_1+ t_2)$, where $v_{\text{PP}}$ is a linear combination of the velocities of the various WPs~\cite{suppmat}.
    In this way we obtain a divergent contribution to $\chi^{(3)}_{\text{PP};d}$ of the form (\ref{eq:pump-probe-scaling}) with $\mathcal{C}_{\text{PP}} = 4 \pi \sum_{n, \text{odd}} \mathcal{C}_{\text{PP}}^{(n)}$, where
	\begin{equation}
	\label{eq:mathcal-C-PP-n}
	   \! \mathcal{C}_{\text{PP}}^{(n)} = -\frac{|{F}_1(0)|^2}{n!}\!\!\int\!\!\! \frac{d^n \vec{k}}{(2\pi)^{n}} \delta\Big(\!\sum_{j=1}^n k_j\Big)\! \left|{F}_n(\vec{k})\right|^2 v_{\text{PP}}(\vec{k}).\!\!\!
	\end{equation}
	Note that for $g\gtrsim 1.1$, $\mathcal{C}_{\text{PP}}$ is dominated by  $\mathcal{C}_{\text{PP}}^{(3)}$, with higher values of $n$ yielding numerically smaller corrections. This allows us to numerically estimate $\mathcal{C}_{\text{PP}}$, reported in Fig.~\ref{fig:fig2} for representative values of $g$.

    Other types of processes only give rise to subleading contributions at late times. Scattering-connected processes where $M( t_1)$ creates more than one QP subsequently annihilated by $M( t_1+ t_2)$  give contributions  suppressed as $ t_2\to\infty$. This follows since QPs created by $M( t_1)$  spread ballistically from one another and therefore cannot be annihilated by a single local operator. For processes that are not simply scattering-connected, the position where all operators act is fixed by the ballistic propagation of the various QPs, therefore we expect their contributions to remain finite at late times.

	\textit{Form-factor expansion.---}%
    We bolster the semiclassical result with an exact calculation of the late-time asymptotics of $\chi_{\text{PP};d}^{(3)}$ via a form-factor expansion~\cite{PhysRevB.78.100403,EK09,pozsgay2008form,pozsgay2010form,PhysRevLett.106.227203,Calabrese_2012,Schuricht_2012,PhysRevLett.109.247206,10.21468/SciPostPhys.9.3.033}.
    The main steps are as follows. As a result of integrability the exact energy eigenstates $|\boldsymbol{p}_N\rangle\equiv|p_1,\dots p_N\rangle$ can be labelled by the momenta of the quasiparticles. Inserting resolutions of the identity in terms of energy eigenstates between each pair of operators yields
\begin{equation}
 \chi_{\text{PP};d}^{(3)} = \sum_{l,m,n\geq 0} \chi_{\text{PP};d}^{(3), [l,m,n]}(t;\boldsymbol{K}_{n_a},\boldsymbol{p}_{n_b},\boldsymbol{k}_{n_c})\ ,
\end{equation}
    where $\chi_{\text{PP};d}^{(3), [l,m,n]}\propto\braket{0|M|\vec{K}}\!\braket{\vec{K}|M|\vec{p}}\!\braket{\vec{p}|M|\vec{k}}\!\braket{\vec{k}|M|0}$. A key property of these matrix elements~\cite{Bugrii2001, BUGRIJ2003390, Gehlen_2008, Iorgov_2011} is the presence of kinematic poles: as $p_i$ approaches $k_j$ the matrix element becomes singular, $\braket{\vec{p}|M|\vec{k}} \sim \frac{1}{p_i-k_j}$. These contributions ultimately give rise to the late-time divergence in $ \chi_{\text{PP};d}^{(3)}$~\cite{suppmat}.
    
	To benchmark the  semiclassical picture, we consider its simplest  non-zero contribution, from $\mathcal{C}_{\text{PP}}^{(3)}$ (sketched in Fig.~\ref{fig:fig1}). Counting the number of QPs before and after each operator $M$, we expect that $\mathcal{C}_{\text{PP}}^{(3)}$ is linked to $\chi^{(3);[3, 4, 3]}_{\text{PP};d}$ in the form-factor expansion.
	Numerically evaluating  $\chi^{(3);[3, 4, 3]}_{\text{PP};d}$  we see good agreement with the semiclassical expectation (see inset of Fig.~\ref{fig:fig2}).
	
	We can also use the form-factor approach to compute the leading correction to \eqref{eq:pump-probe-scaling}, which scales as $\sqrt{ t_2}$~\cite{BABUJIAN2017122, PhysRevB.94.155156}. We conjecture that this can be interpreted as the effect of scattering-connected processes where $M(0)$ and $M( t_1+ t_2)$ create and annihilate a single QP, which can scatter with the QP being exchanged between $A( t_1)$ and $A( t_1+ t_2)$ when their trajectories are smeared by the broadening of QP wave-packets due to dispersion, omitted in the leading semiclassical analysis.

    \textit{Numerical benchmark in the spin-$1$ AKLT chain.---}%
    To confirm that the semiclassical analysis extends beyond integrable models, we  numerically study the spin-$1$ AKLT chain~\cite{PhysRevLett.59.799, Affleck1988}. The model is not integrable, and it features a stable QP mode in the range of momenta $[\tilde{q}, 2\pi-\tilde{q}]$ with $\tilde{q}\simeq 0.4 \pi$~\cite{PhysRevB.102.014447}. Given that $q=0$ does not feature any stable QP mode, we study a finite-momentum response function
    $
    \chi_{\text{PP};d}^{(3)}(q_1,q_2;t_1,t_2) = \frac{2}{L}
    \Im\bra{0}M(-q_1,0)M( -q_2,t_1+ t_2) M(q_2, t_1) M(q_1,0)\ket{0}_C,
    $
    where $M(q,t) =\sum_j e^{iqj} S_j^z(t)$ and $S_j^z$ denotes the spin-$1$ operator along the $z$-axis acting on site $j$.

\begin{figure}
        \centering
        \includegraphics[width=\linewidth]{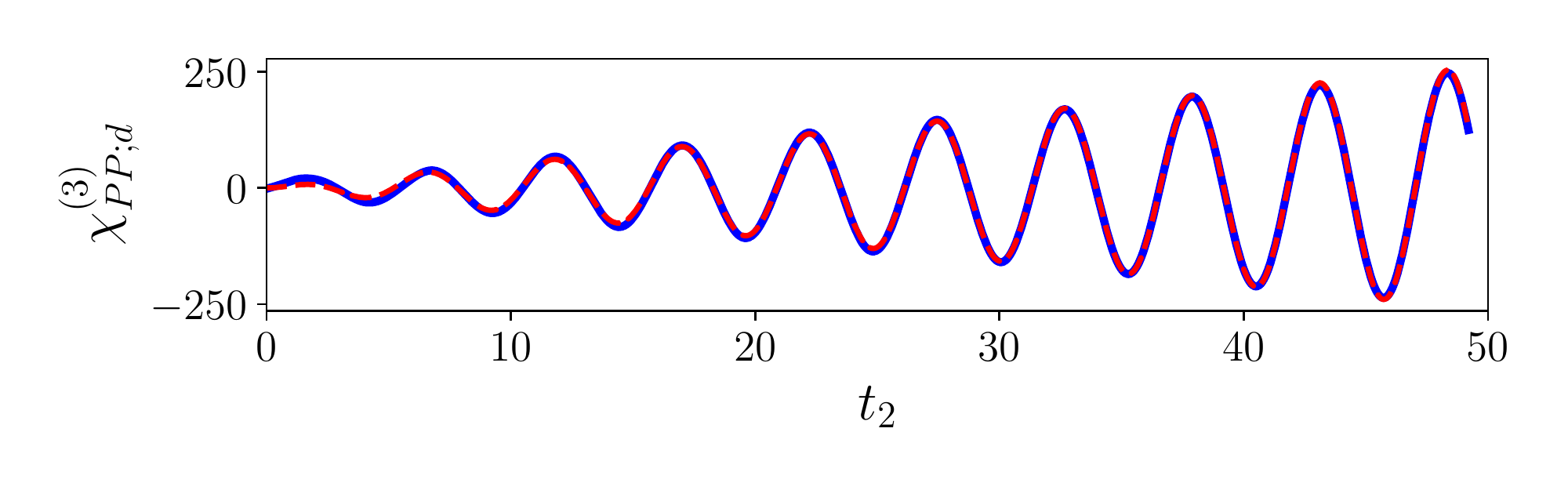}
        \caption{$\chi_{\text{PP};d}^{(3)}(q_1,q_2)$ (blue line), for $t_1=0$, $q_1=\pi$, and $q_2=2\pi/3$. The data is obtained through tDMRG simulation of the spin-$1$ AKLT chain. A red dashed line indicate a fit of the form $A t_2 \sin(\eps(q_2)t_2-\phi)$ consistent with our wave-packet analysis.}
        \label{fig:fig3}
    \end{figure}

    We have computed $\chi_{\text{PP};d}^{(3)}(q_1,q_2;t_1,t_2)$ using tDMRG. The results for $t_1=0$, $q_1=\pi$, and $q_2=2\pi/3$ are reported in Fig.~\ref{fig:fig3}. At sufficiently late times $\chi_{\text{PP};d}^{(3)}$ is well-fitted by the functional form $A t_2 \sin(\eps(q_2)t_2-\phi)$, which is consistent with our wave-packet analysis.
    Here $A$ and $\phi$ are fitting parameters, while $\eps(q_2)$ is independently determined from the numerical computation of the two-point function.

    \textit{Discussion.---}%
    We have shown that in the transverse-field Ising model the nonlinear pump-probe response is characterised by a linear-in-time divergence. We will now argue that this is in fact a general feature of any interacting translationally-invariant many-body system that has a stable, gapped single-particle excitation. Such a behaviour is suggested by our analysis where we demonstrated that the late-time pump-probe signal $\Xi_{\text{PP}}$ can be re-expressed as the difference of two-point correlation functions at zero and finite temperatures and hence is $O(1)$; the divergent response reconciles this result with the perturbative expansion in powers of the applied fields. We therefore expect this linear-in-time growth to be quite general and as we have argued, in many cases visible also on intermediate timescales at low finite temperature. 
    We have developed a semiclassical picture of WP propagation and scattering that identifies the processes that give rise to the divergence.
    Our discussion has focused on the simple case of the Ising model; however, the form factor calculations in fact partially generalizes to other integrable theories as is shown in a forthcoming work~\cite{the_other}.
    More importantly, the semiclassical arguments generalize to non-integrable models, and even to finite temperature~\cite{suppmat}.

    An enticing possibility suggested by our work is the measurement of scattering matrices from third-order response functions. In $\chi_{\text{PP};d}^{(3)}(q_1,q_2;t_1,t_2)$, if the $n=1$ contribution is the dominant one --- as can e.g. be achieved  using a frequency-modulated pump with negligible amplitude to excite the system at energies greater than $2 \Delta$ --- $\chi_{\text{PP}}^{(3)}$ can be expresses in terms of the scattering matrix $S(q_1,q_2)$ and data that can be extracted from linear response, allowing us to read off $S(|q|,0)$ from the divergent piece of $\chi_{\text{PP}}^{(3)}$.~\cite{suppmat}

    Finally, it would be interesting to understand if similar late-time divergences emerge in $d>1$. The non-perturbative argument is evidently independent of dimension, suggesting that this is indeed the case.
    We leave a more detailed investigation of this intriguing possibility to future work. It would also be interesting to understand if these or similar mechanisms are responsible for the anomalously large nonlinear response  recently observed in 2DCS experiments~\cite{2022arXiv220404203C}.

\begin{acknowledgments}
\textit{Acknowledgments.---}%
We thank Abhishodh Prakash, Nick Bultinck and especially Sounak Biswas for many insightful discussions. We also thank Sounak Biswas and Max McGinley for collaboration on related projects and Max McGinley for useful comments on the manuscript. MPS simulations have been performed using the ITensor library~\cite{itensor, itensor-r0.3}. We acknowledge support from the  European Research Council under the European Union Horizon 2020 Research and Innovation Programme, Grant Agreement No. 804213-TMCS (M.F., S.A.P.), the UK Engineering and Physical Sciences Research Council via Grant No. EP/S020527/1 (F.H.L.E.), the US National Science Foundation under Award No. DMR-1653271 (S.G.), the US Department of Energy, Office of Science, Basic Energy Sciences, under Early Career Award No. DE-SC0019168 (R.V.), and the Alfred P. Sloan Foundation through a Sloan Research Fellowship (R.V.).
Statement of compliance with EPSRC policy framework on research data: This publication is theoretical work that does not require supporting research data.
\end{acknowledgments}

\end{document}


\title{Supplementary material for \\
        ``Divergent nonlinear response from quasiparticle interactions''}
	\date{\today}
	\author{Michele Fava}
	\affiliation{Rudolf Peierls Centre for Theoretical Physics,  Clarendon Laboratory, Oxford OX1 3PU, UK}
    \affiliation{Philippe Meyer Institute, Physics Department, \'Ecole Normale Sup\'erieure (ENS), Universit\'e PSL, 24 rue Lhomond, F-75231 Paris, France}
	\author{Sarang Gopalakrishnan}
	\affiliation{Department of Physics, The Pennsylvania State University, University Park, PA 16802, USA}
	\affiliation{Department of Electrical and Computer Engineering, Princeton University, Princeton, New Jersey 08544, USA}
	\author{Romain Vasseur}
	\affiliation{Department of Physics, University of Massachusetts, Amherst, MA 01003, USA}	
	\author{Fabian Essler}
	\affiliation{Rudolf Peierls Centre for Theoretical Physics,  Clarendon Laboratory, Oxford OX1 3PU, UK}
	\author{S. A. Parameswaran}
	\affiliation{Rudolf Peierls Centre for Theoretical Physics,  Clarendon Laboratory, Oxford OX1 3PU, UK}
	
	\date{\today}
	
	\maketitle

    \onecolumngrid
    
    \tableofcontents

\section{Pump-probe signal and its perturbative expansion}
\label{sec:pump-probe-perturbation}

The pump-probe setup can be quite generally described by a time-dependent Hamiltonian
\begin{equation}
H(t)=H_0+V_{\text{pump}}(t)+V_{\text{probe}}(t)\ .
\end{equation}
For our purposes it is sufficient to consider
\begin{equation}
\label{eq:perturbation}
V_{\text{pump}}(t)=\mu \ \delta(t)\ M\ ,\qquad
V_{\text{probe}}(t)=\mu'\ f(t)\ M\ ,
\end{equation}
where $f(t)$ is a smooth function of time and we will take $\mu'$ to be a very small parameter. Then we have
\begin{align}
|\psi(t=0^+)\rangle=e^{-i\mu M}|\psi(t=0^-)\rangle\equiv |\psi_\mu\rangle\ .
\end{align}
At later times the quantum state of the system is then
\begin{equation}
\ket{\psi_{\mu,\mu'}(t)} = {T} e^{-i\int_{0+}^t dt'[H_0+\mu'f(t')M]}\ket{\psi_\mu},
\end{equation}
with $T$ indicating that the exponential is time-ordered.
Defining expectation values
\begin{align}
\langle{\cal O}\rangle_{t,\mu,\mu'}\equiv\langle\psi_{\mu,\mu'}(t)| {\cal O}|\psi_{\mu,\mu'}(t)\rangle,
\end{align}
one then has
\begin{align}
\frac{1}{L}\left[\langle M\rangle_{t,\mu,\mu'}-\langle M\rangle_{t,\mu,0}\right]=
\mu'\int_{0^+}^t dt'\ f(t')\ \chi_\mu^{(1)}(t,t')\nonumber\\
+(\mu')^2\int_{0^+}^t dt'\int_{0^+}^{t'}dt'' f(t')f(t'')\ \chi^{(2)}_{\mu}(t,t',t'')
+{\cal O}({\mu'}^3),
\label{nonlinearresponse}
\end{align}
where the response functions are given by
\begin{align}
\chi_\mu^{(1)}(t,t')&=-\frac{i}{L}
\langle\psi_\mu|[M(t),M(t')]|\psi_\mu\rangle
\ ,\quad M(t)=e^{iH_0t}Me^{-iH_0t}\ ,\nonumber\\
\chi_\mu^{(2)}(t,t',t'')&=-\frac{1}{L}\langle\psi_\mu|[[M(t),M(t')],M(t'')]|\psi_\mu\rangle
\ .
\end{align}
 Note that we have chosen to normalize every response function by $1/L$, so that it remains finite in the $L\to\infty$ limit. Finally, we imagine repeating the linear response measurement with $\mu=0$ and subtracting the two results to obtain
\begin{equation}
\Xi_{\text{PP}}(t,t') = \chi^{(1)}_\mu(t,t') - \chi^{(1)}_0(t,t').
\end{equation}
This is the form of the pump-probe response given in eqn (1) of the main text.
Choosing our initial state before the pump process to be the ground state $|\psi(t=0^-)\rangle=|0\rangle$ and then expanding $\Xi_{\text{PP}}(t,t')$ to second order in $\mu$ we finally arrive at the definition of $\chi^{(3)}_{\text{PP};d}$ given in Eq.~(2)
\begin{align}
\Xi_{\text{PP}}(t_1+t_2,t_1)  &= -i\frac{\mu^2}{L} \langle 0|M(0) [M( t_1+ t_2), M( t_1)] M(0)|0\rangle \nonumber\\
        &\quad + i \frac{\mu^2}{2L} \langle 0|\left\{\left(M(0)\right)^2,  [M( t_1+ t_2), M( t_1)]\right\}|0\rangle + O(\mu^3)\nonumber\\
        &=\mu^2\left[\chi^{(3)}_{\text{PP},d}(t_1+t_2,t_1)+\chi^{(3)}_{\text{PP},c}(t_1+t_2,t_1)\right]+{\cal O}(\mu^3).
\end{align}
Here $\{\cdot,\cdot\}$ denotes an anticommutator and in the last line we have used that

 \begin{align}
\chi^{(3)}_{\text{PP}; d}(t,t')&=  -\frac{i}{L} \Braket{0|M(0) [M( t), M( t')] M(0)|0}  - 2 \Braket{0|M(0) M(0)|0} \Im \Braket{0|M( t) M( t')|0}\ ,\nonumber\\
\chi^{(3)}_{\text{PP}; c}(t,t')&=\frac{i}{2L}\Braket{0|\left\{\left(M(0)\right)^2,  [M( t), M( t')]\right\}|0}  + 2 \Braket{0|M(0) M(0)|0} \Im \Braket{0|M( t) M( t')|0}\ .
    \end{align}

\section{Long-time divergences and non-perturbative effects in integrable and non-integrable systems}

    In the main text, we argued that the late-time divergence of $\chi_{\text{PP};d}^{(3)}$ can be understood  on general grounds, by noting that at late times the perturbation of strength $\mu$ has $O(\mu^0)$ effects.
    While in the main text we framed our arguments for non-integrable systems, in this section we further elaborate on the connection between long-time divergences and non-perturbative effects in integrable systems, where the argument can be presented in a more controlled fashion.
    
    As in the main text, we work in the limit of large $t_1$, when we can assume that the density matrix can  be approximated by a generalized Gibbs ensemble (GGE), i.e. the density matrix takes the  form $\rho\sim\exp(-\sum_k \lambda_k n_k)$, obtained by maximizing entropy subject to the conservation of the number of QPs $n_k$ in mode $k$.
    Here the $\lambda_k$ could in principle be fixed by imposing $\Tr(\rho\, n_k) = \braket{0|e^{i\mu M} n_k e^{-i\mu M}|0}$. 
    The first line of Eq.~(1) is then a linear-response function on top of $\rho$. Given that $\rho$ is a finite energy-density state, this is known to decay exponentially~\cite{10.21468/SciPostPhys.9.3.033}, viz.
    \begin{equation}\label{eq:PPcorr}
        \langle e^{i\mu M(0)} [M( t_1+ t_2), M( t_1)] e^{-i\mu M(0)}\rangle\sim e^{-\gamma_\mu  t_2},
    \end{equation}
    with $\gamma_\mu$ determined by the set $\{\lambda_k\}$.
    Thus $\Xi_{\rm PP}$ becomes a difference of correlators at finite and zero energy density, and as the latter exhibits undamped oscillations in time we have $\Xi_{\text{PP}}\sim O(\mu^0)$. On the other hand, expanding $\Xi_{\rm PP}$ in a power series in $\mu$ (at large, finite $L$) gives $\Xi_{\text{PP}}=\sum_{n\geq 2}\mu^{2n} g_{2n}( t_1, t_2).$
    This can only become ${O}(\mu^0)$ at late times $ t_{1,2}$ if the limits of $\mu\to 0$ and $t_{1,2}\to\infty$ do not commute and the expansion is not uniformly convergent. A natural possibility is to have an ``infrared divergence'' that needs to be summed to all orders in $\mu$. In the Ising case we expect exponentiation of the dominant contribution in the late-time regime, so that
\begin{equation}
\Xi_{\text{PP}}(t_1+t_1,t_1)\sim (1-e^{-c\mu^2 t_2})\langle 0|[M( t_2)M(0)]|0\rangle\sim\begin{cases}
c\mu^2t_2\langle 0|[M( t_2)M(0)]|0\rangle & \text{ if } \mu^2 t_2\ll 1\ ,\\
e^{-c\mu^2 t_2}\langle 0|[M( t_2)M(0)]|0\rangle & \text{ if } \mu^2 t_2\gg 1.
\end{cases}
\end{equation}
These considerations suggest that $\chi^{(3)}_\text{PP}$ diverges whenever we have a stable QP excitation at zero momentum, which gets damped at finite energy densities. Thus the nonlinear response initially shows a linear divergence, probed in the response limit, that eventually saturates at late times $ t\gg1/\mu^2$.

\section{Explicit form of \sfix{$v_{\rm PP}$}}

    \begin{figure}
        \centering
        \includegraphics{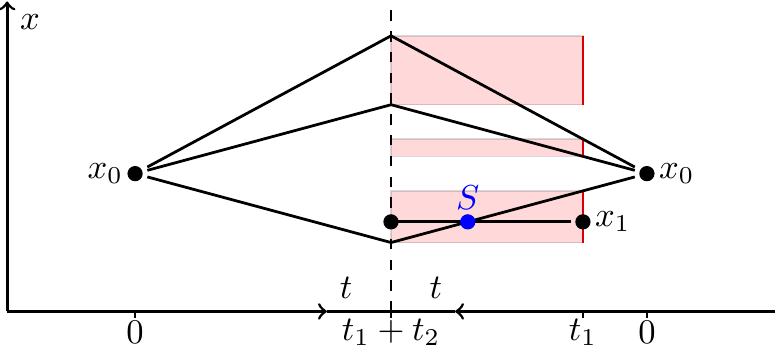}
        \caption{
        [Reproduced from Fig.~1 of the main text.]
        Cartoon of processes contributing to the late-time divergence of the third-order response $\chi^{(3)}_{PP;d}\sim\langle M(0)  M( t_1+ t_2) M( t_1) M(0)\rangle$. A dashed line, corresponding to time  $ t= t_1+ t_2$ separates the bra (left) and ket (right) sides of the time evolution; time increases towards this line.
        Solid lines denote QP worldlines, and circles denote the action of the operator $M$. When two lines cross, the amplitude for the diagram is multiplied by $S=-1$. For a fixed $x_0$, the red segments indicate the set of $x_1$ giving rise to a scattering-connected contribution. The length of the red set is proportional to the overall timescale, leading to the linear divergence of $\chi^{(3)}_{{\rm PP};d}$. }
        \label{fig:fig1-reproduction}
    \end{figure}
    
	Referring to the process depicted in Fig. 1 of the main text (also reproduce in Fig.~\ref{fig:fig1-reproduction} for the reader's convenience) the set of $x_1$ leading to scattering-connected processes has length $v_{\text{PP}}( t_1+ t_2)$, where $v_{\text{PP}}$ has the dimension of a    
    velocity.
    We give here an explicit formula for $v_{\rm PP}$ in terms of the velocities of the quasiparticles. Denoting with $k_1,\dots,k_n$ the momenta of the WPs produced by the pump, $v_{\rm PP}$ is obtained as follows: we define the set of positions $Y=\{v(k_1) t_1,\dots,v(k_n) t_1,v(k_1)( t_1+ t_2),\dots, v(k_n)( t_1+ t_2)\}$ and then reorder the elements $y_j\in Y$ such that $y_1<y_2<\dots<y_{2n}$. Then
    $v_{\text{PP}}(\vec{k}) = \frac{1}{ t_1+ t_2} \sum_{j=1}^{2n} (-1)^j y_j$.

\section{Extension to Two-Dimensional Coherent Spectroscopy (2DCS)}%
\label{sec:2DCS}

    While the main text focuses on pump-probe spectroscopy, similar conclusions apply to 2DCS experiments, where the system is perturbed by a pair of field pulses at varying times.
    More explicitly, in Eq.~\eqref{eq:perturbation} for 2DCS we have $f(t)=\delta(t-t_1)$ and $\mu'$ is comparable to $\mu$. By performing the experiment several times, one can measure 
    \begin{equation}
        \Xi_{\rm 2DCS} = \frac{1}{L}\left[\langle M\rangle_{t,\mu,\mu'}-\langle M\rangle_{t,\mu,0} - \langle M\rangle_{t,0,\mu'} + \langle M\rangle_{t,0,0}\right].
    \end{equation}
    We can then proceed to expand $\Xi_{\rm 2DCS}$ perturbatively in $\mu$ and $\mu'$ up to third order
    \begin{align}
        \Xi_{\rm 2DCS} &=  \mu'\mu^2 \left[\chi^{(3)}_{\text{PP},d}(t_1+t_2,t_1)+\chi^{(3)}_{\text{PP},c}(t_1+t_2,t_1)\right] \nonumber\\
        &+ (\mu')^2 \mu \left[\chi^{(3)}_{\text{NR},d}(t_1+t_2,t_1)+\chi^{(3)}_{\text{NR},c}(t_1+t_2,t_1)\right] + \dots,
    \end{align}
    with ``NR'' standing for ``non-rephasing'' (as motivated by the frequency pattern of the divergence, cf. Eq.~\eqref{eq:NR-contribution} and subsequent discussion)
    \begin{align}
        \chi^{(3)}_{\text{NR}; d}(t,t')&= 
        -\frac{i}{L} \Braket{0|[M(t') M( t) M( t'), M(0)]|0} 
        - 2 \Braket{0|M(t') M(t')|0} \Im \Braket{0|M( t) M(0)|0}\ ,\nonumber\\
        \chi^{(3)}_{\text{NR}; c}(t,t')&=
        \frac{i}{2L}\Braket{0|\left[\left\{M( t), (M( t'))^2\right\}, M(0)\right]|0}  
        + 2 \Braket{0|M(t') M(t')|0} \Im \Braket{0|M( t) M(0)|0}\ .
    \end{align}
    These terms arise from the second order term in (\ref{nonlinearresponse}).
    The nonlinear susceptibilities $\chi^{(3)}_{\rm NR; c}$ and $\chi^{(3)}_{\rm NR; d}$ have an almost identical structure to $\chi^{(3)}_{\rm PP; c}$ and $\chi^{(3)}_{\rm PP; d}$ respectively, with the only difference that the role of $t'$ and $t=0$ have been exchanged.
    Also in this case we have that $\chi^{(3)}_{\rm NR;c}$ is subleading at long times, while $\chi^{(3)}_{\rm NR; d}$ diverges. Its leading processes are those where a shower of $n$ QPs are exchanged between the two copies of $M( t_1)$, and a single QP is exchanged between $M(0)$ and $M( t_1+ t_2)$. 
    Proceeding similarly to the pump-probe case, we find 
    \begin{equation}
    \label{eq:NR-scaling}
        \chi_{\text{NR};d}^{(3)} \simeq 2 t_2 v_{\text{NR}} \sin[\Delta( t_1+ t_2)]  ,
    \end{equation}
    where $v_{\text{NR}}=\mathcal{C}_{\text{PP}}(0)$ has the dimension of a velocity.

    In 2DCS experiments $\Xi_{\rm 2DCS}$, usually analysed in a `two-frequency plane' $(\omega_1,\omega_2)$
    \begin{equation}
        \check\Xi_{\rm 2DCS}(\omega_1,\omega_2) = \int_0^{+\infty} dt_1\, \int_0^{+\infty} dt_2\, e^{i(\omega_1 t_1+ \omega_2 t_2)} \Xi_{\rm 2DCS}(t_1,t_2)
    \end{equation}
    In the two-frequency space our findings  translate into pronounced singularities. This is most easily understood for $\chi_{\text{NR};d}^{(3)}$, which would produce a singularity near the point $(\omega_1,\omega_2)=(\Delta,\Delta)$ of the form 
    \begin{equation}
    \label{eq:NR-contribution}
        \check\Xi_{\rm 2DCS}(\omega_1,\omega_2) \sim \left[(\omega_1-\Delta)+i0^+\right]^{-1} \partial_{\omega_2}\left[(\omega_2-\Delta)+i0^+\right]^{-1}, \qquad \text{near }(\omega_1,\omega_2)=(\Delta,\Delta).
    \end{equation}
    We used the acronym ``NR'' to refer to $\chi^{(3)}_{\rm NR; d}$ as the position of the divergence $(\Delta,\Delta)$ is the one normally associated to ``non-rephasing'' signals in 2DCS.

    A similar, albeit  more complex, structure emerges around the point $(\omega_1,\omega_2)=(0,\Delta)$ due to $\chi^{(3)}_{\text{PP};d}$. Nonetheless, we believe the long-time divergences should be most convenient to diagnose directly in real time.

\section{Extension to the low-temperature regime}%
\begin{figure}[t]
	\centering
        \begin{tikzpicture}[x=1.2cm,y=0.8cm]
        	\draw[thick,->] (-0.5,0) -- (-0.5,4.5) node[anchor=north west] {$x$};
        	\draw[thick,->] (-0.5,0) -- (2,0) node[anchor=south west] {$t$};
            \draw[thick,->] (5.5,0) -- (3,0) node[anchor=south east] {$t$};
            \draw[thick] (2,0) -- (3,0);
            \draw[thick, dashed] (2.5,0) -- (2.5,4.5);

        	\node (a) at (0.5,2) {};
        	\node (b) at (4,1.3) {};
        	\node (c) at (2.5,1.3) {};
            \node (d) at (4.5,2) {};

        	\filldraw[black] (a) circle (2pt) node[anchor=east]{$x_0$};
            \filldraw[black] (d) circle (2pt) node[anchor=west]{$x_0$};
        	\filldraw[black] (b) circle (2pt) node[anchor=west]{$x_1$};
        	\filldraw[black] (c) circle (2pt) node[anchor=west]{};

            \draw[thick] (0.5, 0) -- (0.5, -0.1);
            \node at (0.5, -0.3) {$0$};
            \draw[thick] (4.5, 0) -- (4.5, -0.1);
            \node at (4.5, -0.3) {$0$};
            \draw[thick] (2.5, 0) -- (2.5, -0.1);
            \node at (2.5, -0.3) {$ t_1+ t_2$};
            \draw[thick] (4, 0) -- (4, -0.1);
            \node at (4, -0.3) {$ t_1$};

        	\draw[thick] (a) -- ++(2,-1);

        	\draw[thick] (a) -- ++(2,1);

        	\draw[thick] (a) -- ++(2,2);

        	\draw[thick] (d) -- ++(-2,-1);

        	\draw[thick] (d) -- ++(-2,1);

        	\draw[thick] (d) -- ++(-2,2);

            \draw[thick] (b) -- ++(-1.5,0);

        	\filldraw[blue] (3.1,1.3) circle (2pt);

            \draw[thick, dashed, red] (-0.5, 0.2) -- (2.5, 2.) -- (5.5, 0.2);
            \filldraw[orange] (3.4,1.47) circle (2pt);
            \filldraw[red] (3.675,1.3) circle (2pt);
            \filldraw[orange] (1.6,1.47) circle (2pt);

            \draw[thick, dashed, red] (-0.5, 3.2) -- (2.5, 2.7) -- (5.5, 3.2);
            \filldraw[orange] (2.95,2.78) circle (2pt);
            \filldraw[orange] (3.61,2.89) circle (2pt);
            \filldraw[orange] (2.05,2.78) circle (2pt);
            \filldraw[orange] (1.39,2.89) circle (2pt);
	
	    \end{tikzpicture}
        
	    \caption{
	       Cartoon of low temperature semiclassical processes. The thermal state is approximated as an ensemble of WP states representing thermal QPs, which follow the same trajectories on bra and ket sides (dashed red lines). As they propagate, the thermal QPs can scatter against those produces by $M(0)$ (orange dots) and  by $M(t_1)$ (red dot). Each orange dot on the bra side is canceled by its partner on the ket side, and hence they do not affect the amplitude. This reduces the problem in essence  to that solved in Ref.~\onlinecite{sachdev1997low}.
    	}
	    \label{fig:fig4}
    \end{figure}
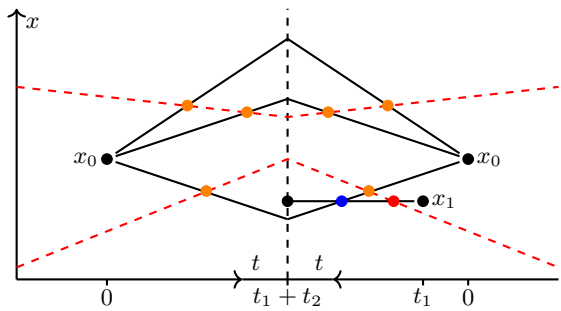

    In the case of the Ising model the semiclassical picture above can be extended to study the low-temperature regime, i.e. $k_B T\ll\Delta$. As in Ref.~\onlinecite{sachdev1997low}, we approximate the thermal density matrix as a diagonal ensemble of WP states, i.e. WPs representing  thermal QPs run on both bra and ket branches with identical trajectories, and affect the response solely via  their purely elastic scattering with the QPs created by the $M$ operators. As depicted in Fig.~\ref{fig:fig4},  thermal QPs will scatter both against those produced by $M(t_1)$ (red dot), and those produced by the pump (orange dots). Crucially, since the  trajectories of both thermal  and pumped QPs  are identical on  bra and ket sides, the latter scattering processes always cancel out and hence do not alter the correlator.
    In contrast, each time a thermal QP scatters with a QP singularly produced at $x_1$ the amplitude of the process acquires a factor $S=-1$. 
    Since the amplitude now fluctuates depending on the number of thermal QPs encountered, averaging over the distribution of the latter results in an exponential decay of the signal controlled by the dephasing rate~\cite{sachdev1997low}
    $\gamma_T \simeq \frac{2 k_B T}{\pi} e^{-\Delta/k_B T}$.
    Therefore the finite-$T$ pump-probe and non-rephasing signals are asymptotically given by
    \begin{align}
        \chi^{(3)}_{\text{PP}}(T>0) &\simeq \chi^{(3)}_{\text{PP}}(T=0) e^{-\gamma_T  t_2}\nonumber\\
        \chi^{(3)}_{\text{NR}}(T>0) &\simeq \chi^{(3)}_{\text{NR}}(T=0) e^{-\gamma_T ( t_1+ t_2)}.
    \end{align}
    [A possibly more accurate estimate might be to replace $|F_1(0)|^2\sin(\Delta \dots)$ with the finite-temperature two-point function \cite{10.21468/SciPostPhys.9.3.033, PhysRevB.78.100403, EK09,pozsgay2008form,pozsgay2010form, DelVecchioDelVecchio2022}.]
    The divergence is thus cut off at  $ t^* \sim e^{\Delta/k_BT}$, and  is observable if $k_B T\lesssim \Delta$.

\section{Semiclassical picture as wave-packet kinematics}

    In this and the next subsection we explain how the semiclassical picture can be made more precise in terms of wave-packets of QPs. In order to keep the discussion as simple as possible we will restrict ourselves to the case of a single QP species. The extension to more general cases will be discussed in Ref.~\onlinecite{the_other}.

    The starting point is the definition of single-QP wave-packet states  in terms of asymptotic scattering states:
    \begin{align}
        \ket{r,k} &= \int \frac{dp}{2\pi} \tilde{W}(p;r,k) \ket{p},
        &
        \tilde{W}(p;r,k) &= \exp\left(-\ell^2\frac{(p-k)^2}{2}\right) e^{i(k-p)r}.
    \end{align}
    Here $\ell$ is some arbitrary microscopic distance, which we assume to satisfy $\ell\gg \xi$.
    The evolution of a wave-packet state is simply given by
    \begin{equation}
        U(t) \ket{r,k} \simeq e^{-i\eps(k)t} \ket{r+v(k)t,k}
    \end{equation}
    up to corrections which can be neglected by choosing $\ell$ s.t. $\ell\gg \sqrt{\eps''(k) ( t_1+ t_2)}$.

    Similarly, one can define wave-packet states in terms of asymptotic scattering states as
    \begin{align}
        \ket{\vec{r},\vec{k}} &= \int \frac{d^n\vec{p}}{(2\pi)^n} \left(\prod_{j=1}^n \tilde{W}(p_j;r_j,k_j) \right)\ket{\vec{p}}.
    \end{align}
    Their time evolution will generally be complicated, however it simplifies in the asymptotic region, i.e. when $|r_i-r_j|\gg\ell$ $\forall\, i,j$.
    In this case, each $r_j$ coordinate will evolve ballistically with the group velocity $v(k_j)$. Furthermore, whenever two wave-packets come in proximity of each other, they will scatter. In 1D and for systems with only one QP species, this will amount to the wavefunction acquiring an overall extra phase: the scattering matrix.

\section{Semiclassical picture for non-integrable models}

    We now detail how the wave-packet description above can be used to naturally extend the analysis in the main text to non-integrable models.

    The starting point is characterizing the state $M \ket{0}$. In non-integrable systems this state will generally be complicated since $M$ can create multiple excitations in the neighborhood of the same point. After some time $t$, however, we expect asymptotic QP states to emerge.
    In this case we can write
    \begin{equation}
    \label{eq:M-on-vacuum}
        e^{-iH t} M(0)\!\ket{0} \simeq \sum_n \frac{1}{\sqrt{n!}} \int dx_0\, \int \frac{d^n \vec{k}}{(2\pi)^{n}} e^{-i x_0 \sum_j k_j} \mathcal{F}_n(\vec{k}) e^{-i\sum_j \eps(k_j) t} \ket{\vec{r}; \vec{k}}
    \end{equation}
    with $\vec{r}$ specified by $r_j=x_0 + v(k_j) t$. Importantly, for a fixed $\vec{k}$, the integral over $y$ is negligible, unless $\left|\sum_j k_j \right| \lesssim \ell^{-1}$.
    Here $\mathcal{F}$ denotes the probability amplitude to produce a given wave-packet state, and is generally challenging to compute from first principles.
    Note that in this subsection, the ``$\simeq$'' sign between two states should not be intended in the sense that the difference between the two has a small norm, but rather that we are isolating a component of the state that will ultimately give the leading contribution to $\chi^{(3)}_{\text{PP};d}$.

    The validity of the equation above can be argued as follows. In fact, \eqref{eq:M-on-vacuum} is not the most general form for a WP state, which would instead be
    \begin{equation}
        e^{-iH t} M(0)\!\ket{0} \sim \sum_n \frac{1}{\sqrt{n!}} \int d^n\vec{x} \, \int \frac{d^n \vec{k}}{(2\pi)^{n-1}} \mathcal{F}'_n(\vec{k}, \vec{x_j}, t)  \ket{\vec{r}; \vec{k}}
    \end{equation}
    for some function $\mathcal{F}'_n$ and with $r_j=x_j + v(k_j) t$.
    [In describing the time dependence we have assumed that no scattering process is taking place. This is justified at sufficiently long times as the $n$ WPs are moving away from each other.]
    The relation above is to be understood in the sense that both sides coincide when projected onto scattering states of quasi-particles, and is an approximation in non-integrable models as it neglects components of the state not captured by the QP description. Nonetheless, given that the LHS is a low-energy state, we expect this approximation to yield accurate results for most models.
    
    Furthermore, the initial positions $x_j$ must necessarily be close: i.e. $|x_i-x_j|\lesssim \xi_A$ for some microscopic length $\xi_A$.
    However, since we are interested in divergences originating at large lengthscales and timescales, we can choose $\ell\gg\xi_A$, so that we can approximate $x_j\simeq x_0$ for all $j$ and $\mathcal{F}'_n(\vec{k}, \vec{x_j})\simeq \mathcal{F}'_n(\vec{k}, x_0)$.
    Finally, the fact that the left-hand side is invariant under translation by any length $a$ requires the $x_0-$dependence of $\mathcal{F}'_n(\vec{k}, x_0+a)e^{i a \sum_j k_j}=\mathcal{F}'_n(\vec{k}, x_0)$, implying that $\mathcal{F}'_n(\vec{k}, x_0)$ is of the form $ e^{-i x_0 \sum_j k_j} \mathcal{F}_n(\vec{k})$.

    Finally, for the semiclassical analysis of the divergence we also need to use that
    \begin{align}
        &M \ket{\vec{r},\vec{n}} =\int_{\substack{\text{asymptotic}\\\text{region}}} dr' F_1(0) \ket{(\vec{r}, r');(\vec{k}, 0)} \nonumber\\
        &+ \sum_j {F}_1^* (0) \ket{(r_1,{\scriptsize \dots},\xcancel{r_j},{\scriptsize \dots},r_n, r');(k_1,{\scriptsize \dots},\xcancel{k_j},{\scriptsize \dots},k_n, 0)} \nonumber\\
        &+\cdots
    \end{align}
    Here the first term amounts to adding a zero-momentum WP at position $r'$, which is possible if $|r'-r_j|\gg\ell$ for all $j$.
    The second line accounts for processes where $M$ annihilates one of the previously-present WPs, instead. Finally, the $\cdots$ denote more complicated outcomes, e.g., where more then one WP is created or annihilated.

    We can now discuss in detail the bra and ket side of Eq.~(2). The bra side can be simply expanded as in \eqref{eq:M-on-vacuum}
    \begin{align}
         \bra{\psi_4}&=\bra{0}M(0)e^{+iH (t_1+t_2)}\nonumber\\
         &\simeq\sum_n \frac{1}{\sqrt{n!}} \int dx'_0\, \int \frac{d^n \vec{k'}}{(2\pi)^{n}} e^{i x_0' \sum_j k'_j}\nonumber \\ &\qquad\times\mathcal{F}^*_n(\vec{k'}) e^{+i\sum_j \eps(k'_j) (t_1+t_2)} \bra{\vec{r_4}; \vec{k'}}
    \end{align}
    with $r_4=x'_0 + v(k'_j) (t_1+t_2)$.
    The ket side can be built progressively. $\ket{\psi_1}=U(t_1)M\ket{0}$ is given by \eqref{eq:M-on-vacuum}. We are then interested in the scenario where $M$ produces an extra WP at $q=0$ on top of $\ket{\psi_1}$:
    \begin{multline}
        \ket{\psi_2}=M\ket{\psi_1} \simeq\\
        \simeq \sum_n \frac{1}{\sqrt{n!}} F_1(0) \int dx_0\,\int dx_1\, \int \frac{d^n \vec{k}}{(2\pi)^{n-1}} e^{-i x_0 \sum_j k_j} \\
         \times\mathcal{F}_n(\vec{k}) e^{-i\sum_j \eps(k_j) t_1} \ket{(\vec{r},x_1); (\vec{k},0)}+\ldots
    \end{multline}
    The states can then be evolved semiclassically up to time $t_2+t_1$. Along this evolution, any of the WPs produced at $t=0$ crosses the trajectory of the $q=0$ WP produced at $t=t_1$, the overall wavefunction acquires a phase given by the product of the scattering matrices for each crossing. We denote this phase as $\mathfrak{S}(x_0,x_1;\vec{k})$.
    Finally the ket side is given by $\ket{\psi_3}=M U(t_2) \ket{\psi_2}$. Here, as described in the main text, we consider the case where $M$ annihilates the $q=0$ WP, thus producing
    \begin{multline}
        \ket{\psi_3}
        \simeq \sum_n \frac{|\mathcal{F}_1(0)|^2}{\sqrt{n!}}  \int\,dx_1 \int dx_0 \int \frac{d^n \vec{k}}{(2\pi)^{n-1}} e^{-i x_0 \sum_j k_j} \mathcal{F}_n(\vec{k}) e^{-i\sum_j \eps(k_j) (t_1+t_2)} \ket{\vec{r}; \vec{k}}\mathfrak{S}(x_0,x_1;\vec{k})+\ldots,
    \end{multline}
    with $r_j=x_0 + v(k_j) (t_1+t_2)$.

    Finally, we can take the overlap between the bra and ket $\braket{\psi_4|\psi_3}$. Comparing the expressions for the two we see that only the terms with $\left\lVert\vec{k}-\vec{k'}\right\rVert\lesssim \ell^{-1}$ and $|x_0-x_0'|\lesssim \ell$ give a significant overlap.
    This gives an overall view of how to extend the semiclassical picture to the non-integrable case. The crucial point, however, remains that for any values of the other integration variables s.t. $\sum_j k_j\approx 0$ we have
    \begin{equation}
        \int dx_0 \left(\mathfrak{S}(x_0,x_1;\vec{k}) -1\right) \varpropto (t_1+t_2),
    \end{equation}
    meaning that upon rescaling both times by some factor $\lambda$, the integral is also rescaled by $\lambda$.
    These considerations therefore lead us to the scaling form (3) also in non-integrable theories.
    A more detailed analysis will be presented in Ref.~\onlinecite{the_other}.

\section{Details of the tDMRG simulations}

    To compute $\chi^{(3)}_{\text{PP};d}$ numerically we proceed as follows. We consider a finite system of size $L$ and compute its ground state $\ket{0}$ through DMRG. [Alternatively, for the AKLT chain the Matrix-Product-State (MPS) form of the ground state is explicitly known.]
    We then compute the two states
    \begin{align}
        \ket{\psi_1} &= U(t_1+t_2) M(q_1) \ket{0}\\
        \ket{\psi_2} &= e^{i q_1 L/2} 
        U(t_2) M(q_2) U(t_1) S^z_{L/2} \ket{0}
    \end{align}
    by using tDMRG for the time evolution and by encoding $M(q)$ as a Matrix Product Operator (MPO) and using standard algorithms to compress MPO-MPS products.
    Finally, we compute $\chi^{(3)}_{PP;d}$ as
    \begin{align}
    \label{eq:chi3-from-numerics}
        \chi^{(3)}_{\text{PP}} &= 2\left[
            \Im \left[ \braket{\psi_1|M(-q_2)|\psi_2}\right]
            - \tilde{C}_2(q_1,0) \Im C_2(-q_2,q_2,t_2)
        \right]+\nonumber\\
        &-\left[
            \Im \left( C_2^*(q_2,q_1, t_1+t_2) \tilde{C}_2(-q_2,t_1) \right)+
            \Im \left( C_2^*(-q_2,q_1, t_1) \tilde{C}_2(q_2,t_1+t_2) \right)
        \right]
    \end{align}
    where
    \begin{align}
        \tilde{C}_2(k,t) &=e^{iq_1L/2} \braket{0|M(-k,t)S^z_{L/2}(0)|0}\\
        C_2(h,k,t) &= e^{iE_0 t} \bra{0}\left[M(h,0)U(t)M(k,0)\ket{0}\right].
    \end{align}
    
    Here $E_0$ denotes the ground-state energy. We note that while we have used translational invariance to obtain the above form, the numerical computations is carried out for an open chain. This is justified for sufficiently large system sizes.
    The asymmetry between the $\tilde{C}_2$ and $C_2$ is to guarantee that the disconnected component of \eqref{eq:chi3-from-numerics} is properly subtracted.
    [The terms in the second line are $0$ in a system with periodic boundary conditions if $q_1\neq q_2$, but not necessarily in a systems with open boundary conditions. Nonetheless, they are $O(L^0 t_{1,2}^0)$, so they do not play an important role in our analysis.]
    
    In our numerical calculation (Fig.~3) the system size is $L=200$, the maximum bond-dimension is $D=400$, and the Trotter step $\delta t=0.01$, and we checked that our results are converged in bond dimension and Trotter step. Simulations have been performed using the ITensor library~\cite{itensor, itensor-r0.3}.
    
    Finally, we note that other techniques to compute nonlinear response functions using MPS techniques have been recently developed~\cite{https://doi.org/10.48550/arxiv.2209.14070, arxiv.2209.00720}.